 \documentclass[preprint,review,12pt]{elsarticle}
\usepackage[fontsize=14pt]{fontsize}

\usepackage{amssymb}
\usepackage{amsmath}
\usepackage[colorlinks=true,
            linkcolor=black, 
            anchorcolor=black,
            citecolor=black,
            urlcolor=blue
            ]{hyperref}
\begin{document}

\begin{frontmatter}



\title{Performance Study of a  Position-sensitive Plastic Scintillator Detector}

 \author[label1]{Shuaike Lv }
 \author[label2]{Changsheng Dai}
 \author[label1]{Dongdong Hu }
 \author[label3]{Tiancheng Zhong}
 \author[label2]{Weifeng Wu}
 \author[label2]{Xinjian Wang \corref{correspondingauthor}}
 \ead{sawxj@mail.ustc.edu.cn}

 \affiliation[label2]{organization={Applied Physics and Optoelectronic Information Research Center, Chizhou University},
            city={Chizhou},
             postcode={247000},
            country={China}}

\affiliation[label1]{organization={Department of Modern Physics, University of Science and Technology of China (USTC)},
             city={Hefei},
             postcode={230026},
             country={China}}

 \affiliation[label3]{organization={Institute of Computing Innovation, Zhejiang University},
             city={Hangzhou},
            postcode={311200},
            country={China}}

\cortext[correspondingauthor]{Corresponding author}

%% Abstract
\begin{abstract}
For a long time, scintillator detectors have suffered from relatively weak spatial resolution due to various influencing factors. Additionally, the high cost of photomultiplier tubes (PMTs) has limited the widespread adoption of scintillator detectors as position-sensitive detectors in particle and nuclear physics experiments. In recent years, thanks to the rapid development of silicon photomultipliers (SiPMs), their excellent cost-performance ratio has led to a renewed interest in scintillator detectors in particle and nuclear physics. This project provides a detailed discussion of a detector based on scintillators coupled with SiPMs, focusing on how to improve the detector's position accuracy. By developing algorithms based on traditional optical propagation, a position resolution of mm level has been achieved. Furthermore, the introduction of a machine learning CNN algorithm has further enhanced the detector's position resolution.
\end{abstract}



%% Keywords

\end{frontmatter}



\section{Introduction}
The Xi’an Proton Application Facility (XiPAF) is dedicated to simulations of environments containing space radiation\cite{b1,b2}. Precise positioning and real-time monitoring of proton beams play a crucial role in enhancing the accuracy of radiation damage testing. In the experiments at XiPAF, the detector is required to have a spatial resolution of less than 1 mm for beam position measurement. This project proposes a position-sensitive detector based on scintillation technology, capable of achieving high-precision beam position measurements.

Scintillation detectors are widely used in particle physics, nuclear physics, medical imaging, and other fields due to its high performance. They can not only serve as a time-of-flight detector to precisely measure the arrival time of particles\cite{b3,b4} but also act as a calorimeter to determine the energy information of particles\cite{b5,b6}.Limited by the large size and high cost of traditional photomultiplier tubes (PMTs), although previous studies have attempted to extend them to position-sensitive detection\cite{b7,b8}, their spatial resolution capability has not seen significant improvement\cite{b9,b10}.
  
In recent years, the breakthrough developments in silicon photomultiplier (SiPM) technology have brought new opportunities for the innovation of position-sensitive scintillation detectors\cite{b11}. Compared to traditional PMTs, SiPMs offer significant advantages such as a compact structure, high photon detection efficiency (PDE), and low cost, making them particularly suitable for high-density array detectors. Research teams have successfully applied SiPM-coupled scintillator detection systems to imaging systems, achieving excellent imaging results through the integration of scintillators with optical fibers coupled to SiPMs\cite{b12,b13}. However, the approach of slotting the scintillator and coupling it with optical fibers has led to increased complexity in mechanical processing and installation procedures. 

This project designs a novel geometric structure by optimizing the coupling method of SiPMs array with scintillators, improving signal processing techniques, introducing neural network algorithms, and developing novel reconstruction algorithms, resulting in a scintillation detector with high position resolution. This solution ensures high position resolution and excellent timing performance while achieving a more compact detector structure and simplified manufacturing processes. Simulation results demonstrate that this novel detector retains the excellent performance of traditional scintillators while effectively addressing the issues of structural complexity and high costs in existing position-sensitive detectors.

The paper is organized as follows. In Sect.2, we describe the design of the detector, including the plastic scintillator and the SiPMs. In Sect.3, we present a Monte Carlo simulation based on GEANT4 to simulate the energy deposited, photon transmission, and the signal output. Sect.4 describes the performance of the detector. And finally,a summary is given in Sect.5\par 

\section{Detector design}
\subsection{Mechanical Design}
\begin{figure}[!htb]
\includegraphics[width=\hsize]{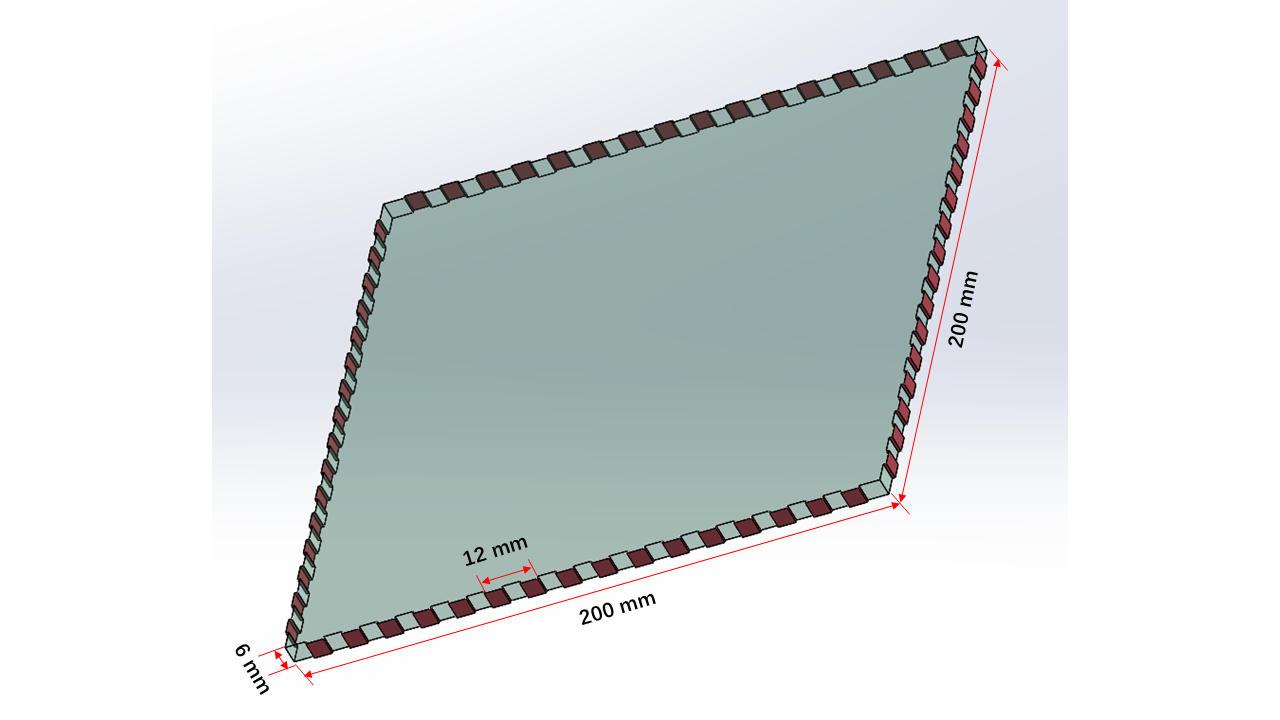}
\caption{Schematic of the plastic scintillator detector with SiPMs array readout}
\label{fig1}
\end{figure}

Fig.~\ref{fig1} illustrates a schematic of the detector structure. The central transparent region represents a rectangular plastic scintillator with an area of 200 mm $\times$ 200 mm and a thickness of 6 mm. The scintillator, an EJ-200 model produced by Eljen Company, is chosen for its superior timing performance—featuring a rise time of 0.9 ns and a decay time of 2.1 ns\cite{b14}. To reduce the number of reflections within the scintillator and thereby enhance timing performance, the scintillator is left unwrapped and exposed directly to air\cite{b15}.On each of the four sides of the scintillator, 16 SiPMs array are uniformly coupled, forming a total of 64 readout channels. The 16 SiPMs on each side are evenly spaced with an interval of 12 mm. In our design, we employ the S13360-6025PE model from Hamamatsu, which features a sensitive area of 6$\times$6 mm$^{2}$, a microcell size of 25 $\mu m$, and a gain coefficient of $10^{6}$, with a quantum efficiency of 40$\%$ at 420 nm\cite{b16}. To further enhance the light collection efficiency, the SiPMs are optically coupled to the plastic scintillator using silicone oil (model EJ500, refractive index n = 1.58).\par 

The operating principle of the detector is as follows: when a particle hit the scintillator, it deposits energy that excites the scintillation material, resulting in the emission of photons. These scintillation photons propagate through the scintillator and reach the SiPMs, where they are converted into electrical signals. By analyzing the timing information from the 64 channels, hit position of the particle can be accurately reconstructed.\par

\subsection{Simulation Study}
\begin{figure}[!htb]
\includegraphics[width=\hsize]{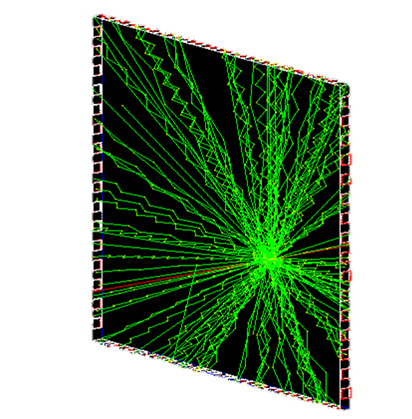}
\caption{Fluorescence transmission in the detector simulated by Geant4}
\label{fig2}
\end{figure}

To optimize the detector's structure and achieve higher spatial resolution, we employed Geant4 for modeling and carried out comprehensive simulations and digitization studies using a minimum ionizing particle (3 GeV/c muon)\cite{b17,b18}.  Fig.~\ref{fig2} illustrates the detailed process in which a muon pass through the scintillator, deposits energy and thereby exciting the scintillator to emit fluorescence. In the figure, the red line indicates the muon's incident direction,while the green curves represent the propagation path of the fluorescence photons within the scintillator. It can be observed that once generated at the muon path, the fluorescence photons are emitted isotropically and, after undergoing multiple reflections, propagate to its lateral surfaces, with a fraction of the photons passing through the silicone oil layer and ultimately hit the SiPMs. In this simulation, the muon was incident uniformly and perpendicular to the surface of the scintillator.

\begin{figure}[!htb]
\includegraphics[width=\hsize]{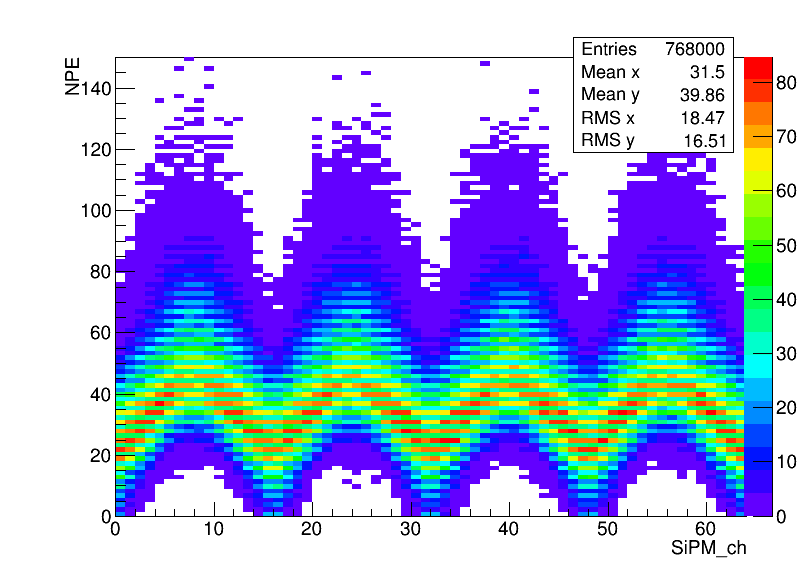}
\caption{Number of photoelectrons generated by the 64 SiPMs array}
\label{fig3}
\end{figure}

Furthermore, photons reaching the SiPMs are converted into photoelectrons via the photoelectric effect. Fig.~\ref{fig3} presents the statistical distribution of the number of photoelectrons(NPE) detected by the SiPMs. The results indicate that there is a significant variation in the NPE received by SiPMs at different locations, and even the same SiPM registers varying NPE across different events. This variability is primarily attributed to differences in the geometric collection efficiency of the SiPMs for photons, as well as the random of the muon incident positions.\par

\begin{figure}[!htb]
\includegraphics[width=\hsize]{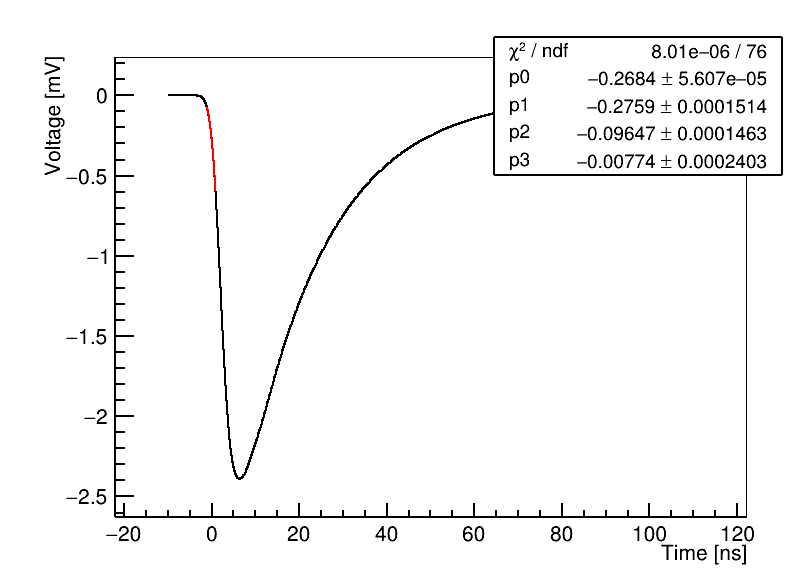}
\caption{Simulated output signal of SiPM}
\label{fig4}
\end{figure}

The signal digitization of the detector is accomplished by reconstructing the output waveform of the SiPM based on the simulated number of photoelectrons (NPE) detected by the SiPM and the arrival time of each individual photoelectron. The single photon–electron (SPE) response waveform $v_{i}(t)$ can be simulated using an experimentally measured SPE signal of the SiPM combined with polynomial fitting. The output signal of the SiPM is the superposition of all the SPE signals in one event, which is given in Relation\cite{b19}:

\begin{equation}
V(t)=\sum_{i=1}^{NPE}v_i\{t-t_{transit}-(t_0)_i\}
\label{equation:V(t)}
\end{equation}

where  $t_{transit}$ represents the transit time of a single photon within the SiPM, which is obtained by sampling from a Gaussian distribution with a mean of 0 and a standard deviation of 20 ps (in the time resolution analysis, the specific value of $t_{transit}$ does not affect the outcome). $t_0$ indicates the arrival time of a single photoelectron at the SiPM. Fig.~\ref{fig4} shows an example of the simulated output waveform from one SiPM when a muon traverses the detector. A constant fraction timing method is applied to the signal, with the timing point defined at the 20\% level of the maximum amplitude on the falling edge, thereby determining the channel's output time. %Subsequently, the timing information from all 64 channels is used to reconstruct the muon's incident position.

\section{Time resolution}
In the process of position reconstruction, a substantial amount of timimg information is utilized, hence we have investigated the time resolution of this detector. In our previous study, the time performance of the scintillator detector with this structure was investigated in detail\cite{b20}. The results demonstrated that although the time performance of an individual channel is significantly influenced by the muon incident position(as illustrated in Fig.~\ref{fig16}), the symmetric arrangement of the SiPMs around the detector allows the averaging of signals across all channels to effectively reduce the timing jitter caused by variations in the muon incidence position. Therefore, the average time from all channels can be used as the arrival time of muons. To eliminate time jitter induced by edge-incident muons, we employed the NPE received by each SiPM as a weighting factor and applied a weighted average algorithm to compute the time:

\begin{figure}[!htb]
\includegraphics[width=\hsize]{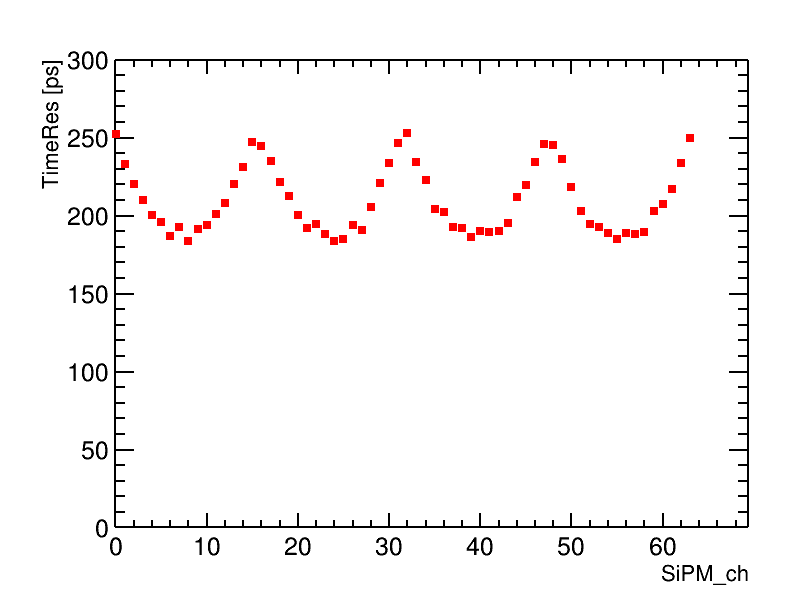}
\caption{The time resolution of different channels.}
\label{fig16}
\end{figure}

\begin{equation}
T_{refcor}= (\sum_{i=1}^{64}N_it_i)/(\sum_{i=1}^{64}N_i)
\label{equation:V(t)}
\end{equation}

where $N_i$ represents NPE of each channel. The resulting time distribution, shown in Fig.~\ref{fig15}, exhibits a single, well-defined peak. A Gaussian fit to this distribution yields a time resolution of 22.29 ps.\par

\begin{figure}[!htb]
\includegraphics[width=\hsize]{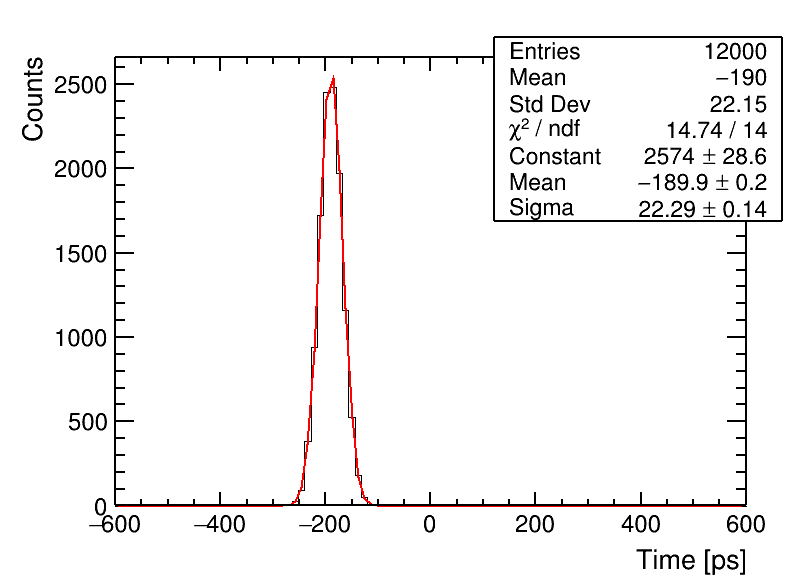}
\caption{The arrival time of muons using weighted average algorithm.}
\label{fig15}
\end{figure}

\section{Spatial resolution}
Utilizing simulated data, we employed two distinct algorithms to investigate the spatial resolution of the detector.The spatial resolution  of the detector is measured in experiments by determining the difference between the reconstructed position of incident muons and their true positions, and the same measurement method is applied in simulations. Therefore, in addition to the inherent spatial resolution of the detector, the ability to more accurately reconstruct the incident position of particles determines the ultimate spatial resolution of the detector. The first method is based on geometrical optics, utilizing a light propagation model to reconstruct the muon incident position, while the second method employs a Convolutional Neural Networks(CNN) algorithm to achieve position reconstruction through neural network techniques.
%\textcolor{yellow}{Firstly, we investigate the spatial performance of the detector. }To reconstruct the muon incident position, two distinct algorithms were employed. The first method is based on geometrical optics, utilizing a light propagation model to reconstruct the muon incident position, while the second method employs a Convolutional Neural Networks(CNN) algorithm to achieve position reconstruction through neural network techniques.

For the convenience of subsequent analysis, this paper defines the geometric center of the detector as the origin of the coordinate system, with the horizontal direction designated as the X-axis and the vertical direction as the Y-axis.\par

\subsection{Photon transmission algorithm}
Assume that in a uniform medium with a refractive index 
n there exists a point light source and a light detector whose volume is negligible. The point light source emits photons isotropically, and the light detector can precisely measure the arrival time t of the photons. According to the principle of straight-line propagation of light, the propagation distance L of the photons can be determined by
\begin{equation}
L = \frac{tc}{n} 
\label{equation:BTV}
\end{equation}

where c is the speed of light in a vacuum. Although this measurement does not directly determine the position of the point light source, it constrains the source to lie on a circle (in the two-dimensional case) centered at the light detector with radius L. If an additional identical light detector is placed at another location in space and the same measurement is performed, a second circle can be obtained. In an ideal scenario, these two circles would be tangent, and their tangency point would correspond to the location of the light source.\par

\begin{figure}[!htb]
\includegraphics[width=\hsize]{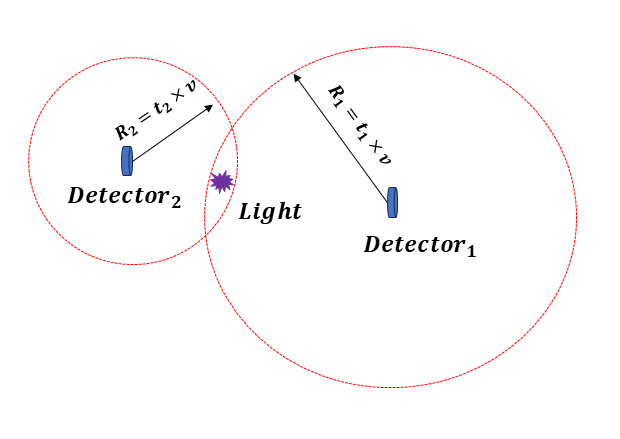}
\caption{Schematic of the photon transmission algorithm.}
\label{fig5}
\end{figure}

The above inference is based on idealized conditions. However, the situation in our detector is more complex. First, photons are emitted randomly along the path of the muon as it traverses the scintillator, with the emission times following an exponential distribution. Second, due to the thinness of the scintillator, only a very small fraction of the photons received by the SiPM are direct; the vast majority undergo multiple reflections within the scintillator. Finally, the SiPM exhibits intrinsic time jitter and its finite volume cannot be neglected. Consequently, the circles derived from any two SiPM measurements will invariably intersect in practical scenarios. Although this precludes an exact determination of the photon emission point, it unequivocally indicates that the emission must occur within the overlapping region of the two circles (see Fig.~\ref{fig5}). This constitutes the basic principle of the light propagation algorithm.\par

\begin{figure}[!htb]
\includegraphics[width=\hsize]{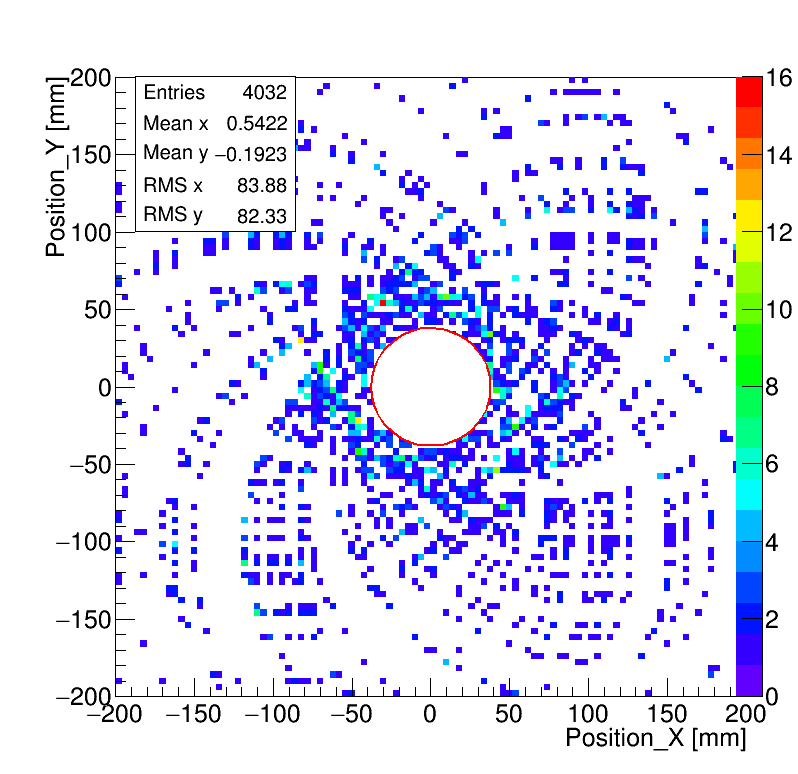}
\caption{Schematic of the Photon transmission algorithm.}
\label{fig6}
\end{figure}

When a muon pass through the detector from the center, the 64 SiPMs provide 64 different timing information. By selecting the geometric centers of any two SiPMs as the centers of circles, and using the timing information from these two SiPMs along with the speed of light in the scintillator (with a refractive index of 1.58), we can draw a set of intersecting circles in the geometric plane. With all 64 SiPMs paired, a total of 2016 distinct intersecting circles can be obtained. The intersection points of all these circles are extracted and plotted on a two-dimensional plane, as shown in Fig.~\ref{fig6}. It can be seen that there is a circular blank region at the center of the plot (marked by the red circle in the figure). The center of this region corresponds to the muon’s incident position.\par

\begin{figure}[!htb]
\includegraphics[width=\hsize]{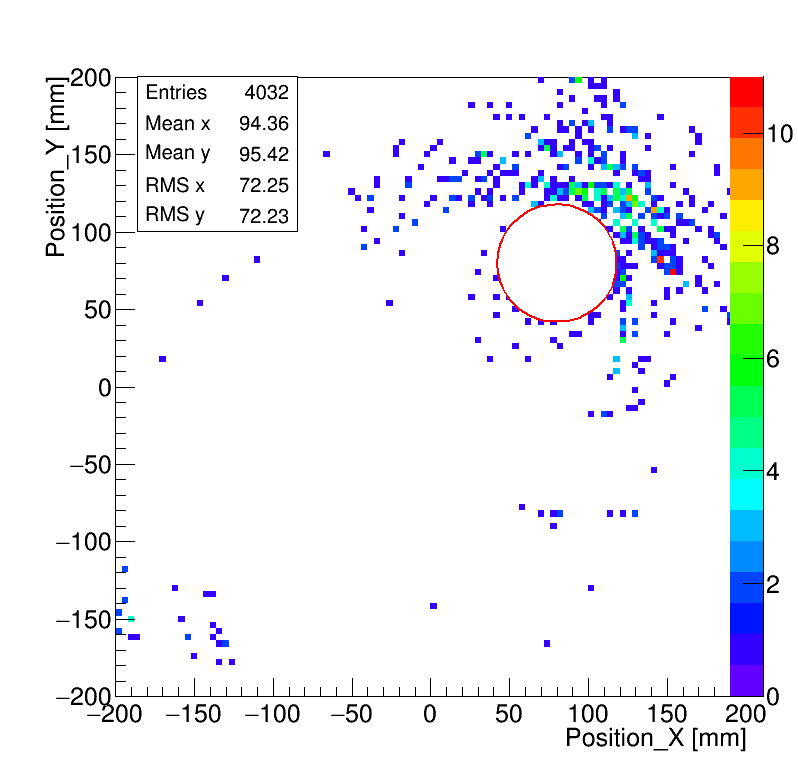}
\caption{Schematic of the Photon transmission algorithm.}
\label{fig7}
\end{figure}

Furthermore, we also investigated the scenario where the muon enters from a position at at (80 mm, 80 mm). The results are shown in Fig.~\ref{fig7}. Compared to the central incidence, the intersection points are more dispersed in this case, but there is still a circular blank region in the upper right corner of the plot. The center of this region again corresponds to the muon’s incident position.

\begin{figure}[!htb]
\includegraphics[width=\hsize]{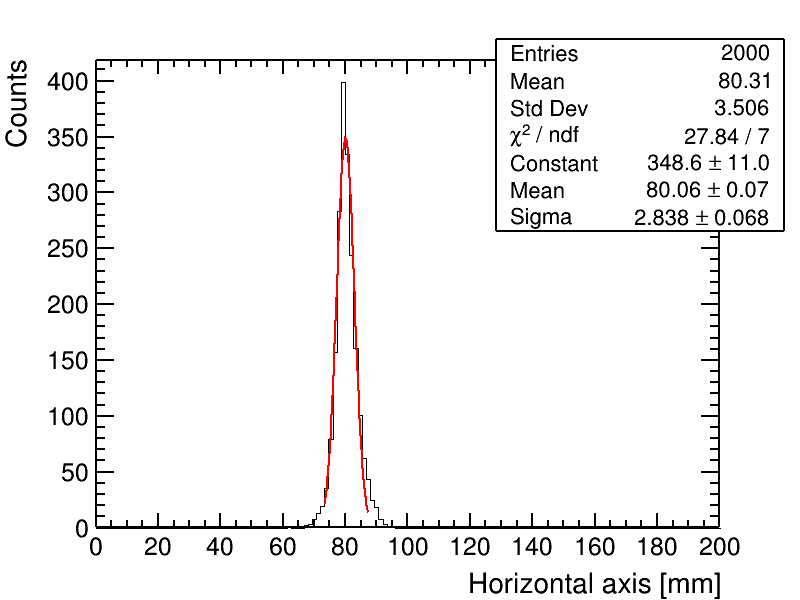}
\caption{Reconstructed X-axis Position at (80 mm, 80 mm).}
\label{fig8}
\end{figure}

\begin{figure}[!htb]
\includegraphics[width=\hsize]{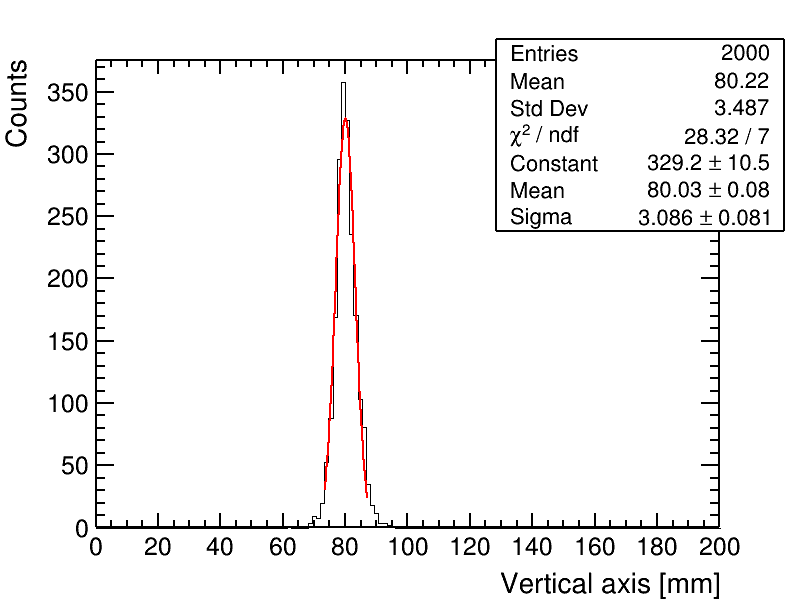}
\caption{Reconstructed Y-axis Position at (80 mm, 80 mm).}
\label{fig9}
\end{figure}

Next, we describe the algorithm used to fit the circular blank region  and determine its center coordinates. Our core idea is to draw the largest possible circle within the blank region while ensuring that the circle does not contain any intersection points. Taking Fig.~\ref{fig7} with as an example, we observe that the intersection points obtained from pairwise combinations of the 64 SiPMs are primarily concentrated near the blank region, which corresponds to the possible incident positions. Therefore, the mean values of the X and Y coordinates of these intersection points from Fig.~\ref{fig7} can be used to assist in determining the center of the circular blank region. Based on this mean value, we further search for the largest possible circle. Specifically, we define a square region centered at the mean value with a certain side length and divide it into a 200×200 grid. By iterating through each grid point and drawing circles, we identify the largest possible circle, and its center is taken as the particle incident position.  Fig.~\ref{fig8} and  Fig.~\ref{fig9} demonstrate the reconstruction performance at the specific incident position of (80 mm, 80 mm) under this algorithm. 

In addition to the off-center incident position at (80 mm, 80 mm), we systematically evaluated the algorithm’s performance across multiple fixed incident positions. The comprehensive results are summarized in Table~\ref{tab1}. The first column of Table~\ref{tab1} lists the true incident positions, while the second and third columns present the reconstructed X- and Y-axis positions obtained using the proposed algorithm. The fourth and fifth columns provide the positioning accuracy along the X- and Y-axis directions, respectively.\par

\begin{table}[h!]
    \centering
    \resizebox{\textwidth/2}{!}{
    \begin{tabular}{|c|c|c|c|c|}  
        \hline
        Position(mm) & MeanX(mm) & MeanY(mm) & SigmaX(mm) & SigmaY(mm)\\  
        \hline
        0, 0 & 0.05 & 0.05 & 2.14 & 2.16 \\ 
        \hline
        20, 20 & 20.03 & 20.24 & 2.17 & 2.21 \\ 
        \hline
        30, 30 & 30.10 & 30.06 & 2.34 & 2.31 \\ 
        \hline
        50, 50 & 50.04 & 50.08 & 2.63 & 2.85 \\ 
        \hline
        80, 80 & 80.06 & 80.03 & 2.83 & 3.08 \\ 
        \hline
    \end{tabular}
    }
    \caption{Reconstruction Results for Fixed Incident Positions}
    \label{tab1}
\end{table}

\begin{figure}[!htb]
\includegraphics[width=\hsize]{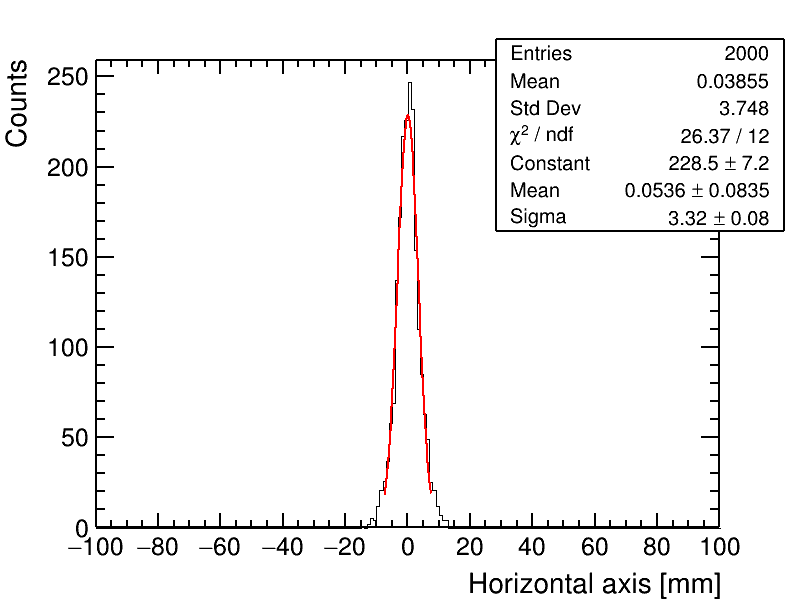}
\caption{Reconstructed X-axis Position under Random Incidence Conditions.}
\label{fig10}
\end{figure}

\begin{figure}[!htb]
\includegraphics[width=\hsize]{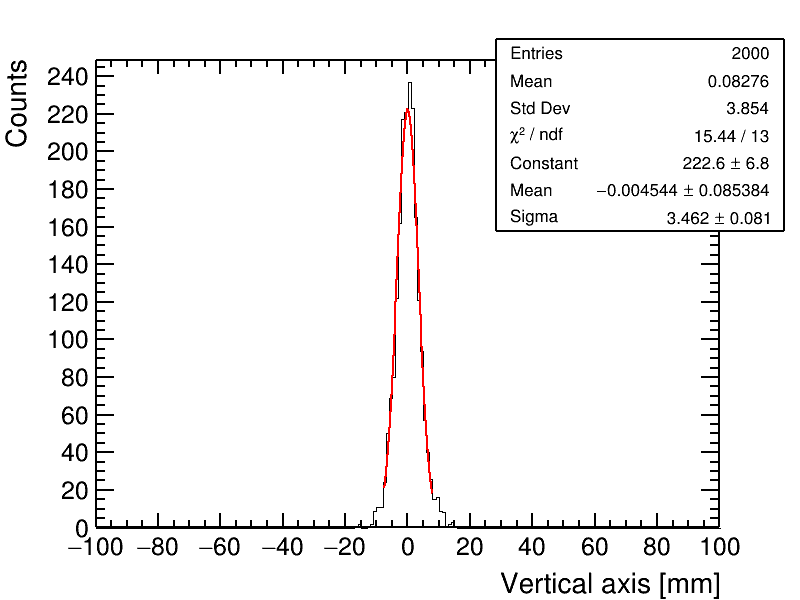}
\caption{Reconstructed Y-axis Position under Random Incidence Conditions.}
\label{fig11}
\end{figure}

To validate the effectiveness of the particle position reconstruction algorithm under random incidence conditions, this study not only investigated multiple fixed incident positions, but also generated uniformly distributed random incident data through Geant4 simulation. The reconstruction performance based on the aforementioned algorithm was analyzed using the simulated data, as shown in Fig.~\ref{fig10} and Fig.~\ref{fig11}. Notably, the reconstructed positions have been subtracted by the true incident coordinates to more clearly demonstrate the reconstruction accuracy.The results show that the position resolution is around 3.4 mm.

\subsection{CNN algorithm}

In particle physics, machine learning has become an important tool for data processing and analysis. It is primarily used for particle identifications\cite{b21,b22,b23}, data representation\cite{b24,b25}, and detector calibration\cite{b26} etc. At the same time, accurately reconstructing the spatial position of incident particles is crucial for physical analysis. Traditional position reconstruction algorithms are typically based on geometric reconstruction or statistical fitting methods, which have certain limitations when dealing with complex detector responses and nonlinear signal features. Machine learning methods, leveraging their powerful nonlinear modeling capabilities, demonstrate significant advantages.Thus, particle incident position reconstruction represents critical application of machine learning\cite{b27}.

In the study, the detector's structure consists of a scintillator surrounded by a uniformly distributed array of silicon photomultipliers (SiPMs), which is spatially symmetric. When incident particles interact with the scintillator, the generated scintillation photons are uniformly distributed within a 4$\pi$ solid angle and propagate along optical paths to reach the SiPM array. Statistically, the SiPMs aligned in the X-Y direction corresponding to the particle's incident position receive the largest number of photons, resulting in the largest signal charge, while adjacent SiPM units exhibit progressively decreasing charge levels. Therefore, for the same incident particle, there is a correlation among adjacent SiPM units on the same side. Additionally, a complementary correlation characteristic exists among the units of the SiPM array on the opposite side. Thus, the spatial information of the incident particle's position is encoded in the time difference information obtained by each SiPM unit, and the response characteristics of these SiPMs can be utilized to infer the particle's incident location. Moreover, due to the symmetric distribution of the SiPM array, the response information of the SiPMs possesses translational invariance, and convolutional neural networks can effectively characterize this symmetric structure. The conventional implementation of the convolution operation in our methodology can be formally expressed as follows\cite{b28}:

\begin{equation}
\mathcal{I}(i,j)=(\mathcal{I}*\mathcal{K})(i,j)
\label{equation:I}
\end{equation}

where $\mathcal{I}$ is the input feature as a 2-dimensional tensor and 2-dimensional tensor $\mathcal{K}$ is used as the convolution kernel. In this paper, both the input features and the convolutional kernels are formulated as 4-dimensional tensors to accurately capture the structural characteristics of the detector system. Let $\mathcal{I} \in \mathbb{R}^{B \times C_{\text{in}} \times H \times W}$ denote the input tensor and $\mathcal{K} \in \mathbb{R}^{C_{\text{out}} \times C_{\text{in}} \times K_{\text{h}} \times K_{\text{w}}}$ represent the convolutional kernel tensor,where $B$ is the batch size, $C_{in}$ and $C_{out}$ are the number of input and output channels respectively, $H$ and $W$ are the height and width of the input feature map, ${K_{\text{h}}}$ and ${K_{\text{w}}}$ are the height and width of the convolutional kernel. The 4D convolution operation at spatial coordinates (${\text{h}}$,${\text{w}}$)is defined as:

\begin{equation}
    \mathcal{S}_{b,c_{\text{out}},{\text{h}},{\text{w}}} = \sum_{c_{\text{in}}} \sum_{k_{\text{h}}} \sum_{k_{\text{w}}} \mathcal{I}_{b,c_{\text{in}},{\text{h}}+k_{\text{h}}-1,{\text{w}}+k_{\text{w}}-1} \cdot \mathcal{K}_{c_{\text{out}},c_{\text{in}},k_{\text{h}},k_{\text{w}}}
    \label{equation:hw}
\end{equation}

The input feature $\mathcal{I}$ is a data set composed of time information
from 64 SiPMs generated by Geant4 simulations. Its dimensionality is
$(batch size,2,16,2)$. Here, the batch size corresponds to the number of samples in each training batch, the value 2 represents the azimuthal classification of the SiPMs (distinguishing between left/right or up/down orientations), 16 denotes the number of combinations of SiPM pair, and 2 corresponds to the timing information of two SiPMs. Additionally, the actual hit position information generated by the Geant4 simulations is used as label attributes. These label attributes record the true hit positions of the particles in the detector, providing a supervised label for model training.

\begin{figure}[!htb]
\includegraphics[width=\hsize]{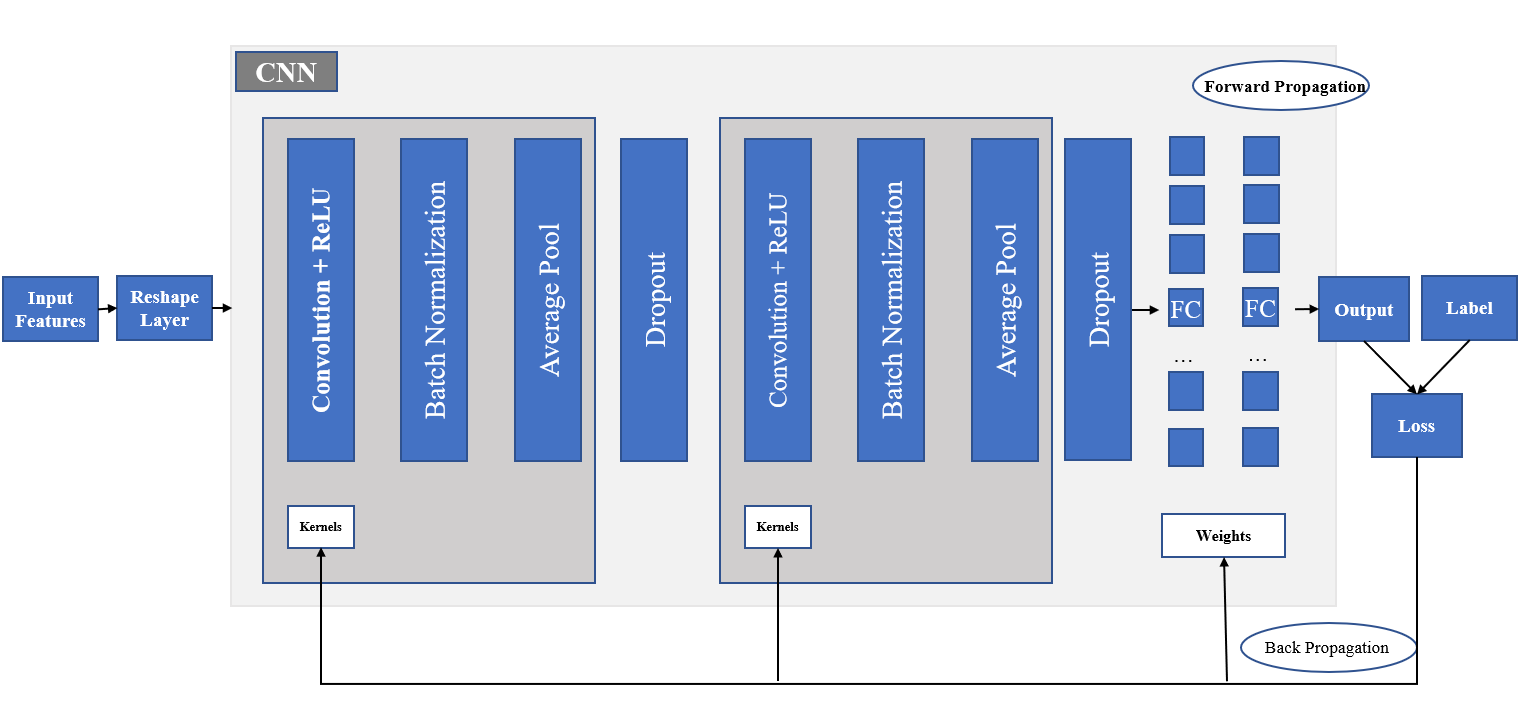}
\caption{An overview of a convolutional neural network (CNN)
 architecture and the training process.}
\label{fig12}
\end{figure}

Finally, the detailed network architecture and data processing pipeline are illustrated in Fig.~\ref{fig12}. We propose a dedicated CNN framework specifically designed for processing two-dimensional tensor data. The input to the model is a 2D tensor with dimensions $(batch size,64)$, which is reshaped into a 4D tensor of shape $(batch size,2,16,2)$. During the feature extraction stage, the model utilizes two convolutional layers, each followed by a Rectified Linear Unit (ReLU) activation function and batch normalization. This design not only accelerates model convergence but also significantly enhances the model's generalization capability\cite{b29}. To further optimize network performance, average pooling layers (AvgPool2d) are integrated between convolutional layers to reduce dimensionality, while dropout layers are incorporated to mitigate overfitting effectively\cite{b30}. Finally, the model flattens the extracted feature maps into a one-dimensional vector through fully connected layers and maps these features to the output space using two linear layers, thereby predicting the 2D coordinates.

In the back propagation process of the  CNN architecture illustrated in Fig.~\ref{fig12},  we employed Mean Squared Error (MSE) as the loss function and utilized the Adaptive Moment Estimation (Adam) algorithm for optimization. Compared with conventional gradient descent methods, Adam demonstrates distinctive advantages: it mitigates parameter update oscillations by integrating current gradients with historical gradient information through exponential moving averages of the first-order moment and second-order moment. The weight update rule of this algorithm is as follows\cite{b31}\cite{b32}:

\begin{equation}
w_{t+1}=w_t+\frac{\lambda}{\sqrt{\Tilde{v}_t}+\epsilon)}*\Tilde{m}_t
\label{equation:wt}
\end{equation}

In the equation, $w_{t}$ denotes the weight parameter at the t-th iteration, $\lambda$ represents the learning rate, and $\epsilon$ is a minimal constant for numerical stability. Here, $m_{t}$ and $v_{t}$ correspond to the exponentially moving average estimates of the gradient's first moment (mean) and second moment (uncentered variance), respectively. Since the algorithm initializes $m_{0}$ and $v_{0}$ to zero vectors, the moment estimates tend to be biased towards zero during initial iterations. To address this bias,we implement a correction strategy to obtain $\Tilde{m}_t$ and $\Tilde{v}_t$: dividing $m_{t}$ and $v_{t}$ by $(1 - \beta_1^t)$ and $(1 - \beta_2^t)$, respectively, where$\beta_1$ and $\beta_2$ are decay rate hyperparameters. This bias correction mechanism effectively eliminates estimation deviations in the initial phase, ensuring stability throughout the training process.

During the model training phase, we employ simulated datasets for optimization while simultaneously monitoring the dynamics of training loss and validation loss. Following an early stopping strategy, the training process is terminated when the validation loss shows no improvement over 2,000 consecutive training epochs, which indicates model convergence. This strategy effectively prevents overfitting through cross-validation mechanisms while ensuring sufficient feature extraction. 

\begin{figure}[!htb]
\includegraphics[width=\hsize]{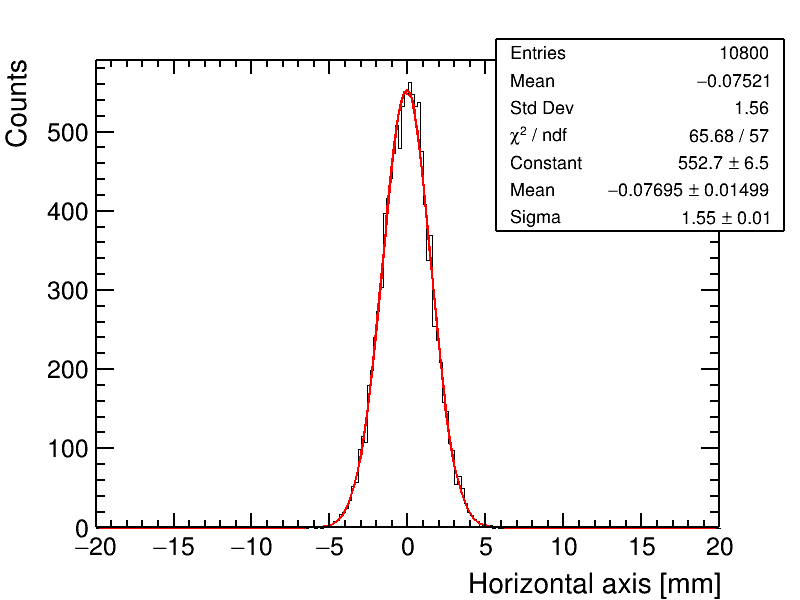}
\caption{Reconstructed X-axis Position under Random Incidence Conditions.}
\label{fig13}
\end{figure}

\begin{figure}[!htb]
\includegraphics[width=\hsize]{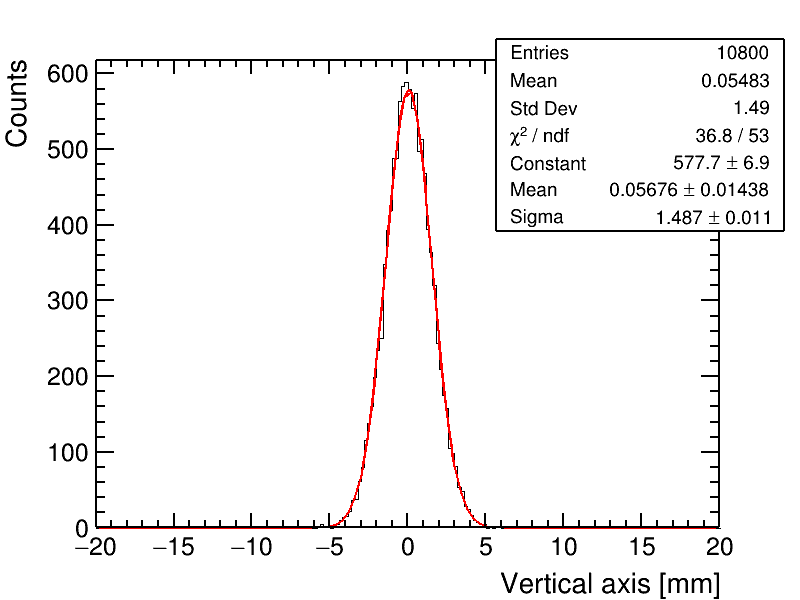}
\caption{Reconstructed Y-axis Position under Random Incidence Conditions.}
\label{fig14}
\end{figure}

The model training and performance evaluation based on the CNN algorithm demonstrate that the target position prediction accuracy is shown in Fig.~\ref{fig13} and Fig.~\ref{fig14}. The results indicate that the position resolution reaches approximately 1.5 millimeters in both the horizontal and vertical directions, demonstrating high positioning accuracy.

\section{Summary}
This paper presents a detector design that integrates a plastic scintillator with SiPMs array, achieving millimeter-level spatial resolution and sub-30 ps timing resolution within a single detector through innovative geometric configurations and advanced algorithms. The detector comprises a bare 200 mm × 200 mm × 6 mm scintillator symmetrically surrounded by 64-channel SiPMs array, optically coupled with silicone oil to maximize light collection efficiency. GEANT4 simulations of 3 GeV/c muon interactions were performed to investigate photon transport dynamics and characterize SiPM signal responses.

For position reconstruction, two complementary algorithms were developed: (1) a geometric optics-based method utilizing photon transport modeling, which localizes incident positions by identifying maximum void circles in intersection plots derived from SiPM time differences. The traditional geometric method achieves a position resolution of approximately 3.5 mm; and (2) a CNN regression model trained on 64-channel timing data, optimized using mean squared error loss and the Adam algorithm with an early stopping strategy to prevent overfitting. The CNN-based algorithm significantly improves this resolution to 1.5 mm. And to optimize timing performance, a photoelectron-weighted averaging algorithm was implemented, effectively mitigating time jitter induced by edge-incident muons and achieving an exceptional time resolution of 22.29 ps.

Furthermore, while the maximal void circle algorithm provides a viable solution, it is not optimal; future work could explore alternative methods such as Hough circle detection to enhance center localization accuracy. The integration of Hough circle detection could provide a more robust and accurate method for center localization, particularly in scenarios with high noise or complex geometries. Additionally, the current CNN algorithm relies solely on timing information; incorporating additional features such as amplitude, photon count, or position data could further improve training performance and reconstruction accuracy. 

Based on these findings, the proposed detector design offers a groundbreaking solution for high-energy physics experiments and particle imaging systems, delivering exceptional temporal sensitivity and spatial precision.

\section{Acknowledgements}
The authors would like to thank Associate Professor Xingyan Chen from the Southwestern University of Finance and Economics for his contributions to CNN reconstruction algorithms. This work was supported by the Chizhou University High level Talent Research Start up Fund [grant number CZ2024YJRC21] and Collaborative Innovation Project of University, Anhui Province[grant number GXXT-2022-088].

\end{document}